\documentclass{lhep}       

\vol{NDM 2020}
\jyear{Egypt}
\pages{Hurghada} 

\received{xx April 2020}
\published{xx April 2020}

\def\be{\begin{equation}}
\def\ee{\end{equation}}
\def\bea{\begin{eqnarray}}
\def\eea{\end{eqnarray}}

\newcommand{\nn}{\nonumber}

\begin{document}

\title{Two component FIMP DM in a $U(1)_{B-L}$ extension of the SM}

\author{Waleed Abdallah,\auno{1,2} Sandhya Choubey,\auno{3} and Sarif Khan\auno{4}}
\address{$^1$Harish-Chandra Research Institute, HBNI, Chhatnag Road,
Jhunsi, Allahabad 211019, India}
\address{$^2$Department of Mathematics, Faculty of Science, Cairo University, Giza 12613, Egypt}
\address{$^3$Department of Physics, School of
Engineering Sciences, KTH Royal Institute of Technology, AlbaNova
University Center, 106 91 Stockholm, Sweden}
\address{$^4$Institut f\"{u}r Theoretische Physik,
Georg-August-Universit\"{a}t G\"{o}ttingen, Friedrich-Hund-Platz 1,
G\"{o}ttingen, D-37077 Germany}

\begin{abstract}
In this work, we discuss two component fermionic FIMP dark matter (DM) in a popular ${B-L}$ extension of the standard model (SM) with inverse seesaw mechanism. Due to the introduced $\mathbb{Z}_{2}$ discrete symmetry, a keV SM gauge singlet fermion is stable and can be a warm DM candidate. Also, this $\mathbb{Z}_{2}$ symmetry helps the lightest right-handed neutrino, with mass of order GeV, to be a long-lived  or stable particle by choosing a corresponding Yukawa coupling to be very small. Firstly, in the absence of a GeV DM component (i.e., without tuning its corresponding Yukawa coupling), we consider only a keV DM as a single component DM produced by the freeze-in mechanism. Secondly, we study a two component FIMP DM scenario and emphasize that the correct ballpark DM relic density bound can be achieved for a wide parameter space.
\end{abstract}

\maketitle


\section{Introduction}
The standard model (SM) is a very successful theory in describing nature. But it can not
explain a number of phenomena - two of the most important ones being the presence
of dark matter (DM) and non-zero tiny neutrino mass.
To address these two issues, we need to extend the
SM particle content and/or its gauge group.
The non-thermal DM production via the so-called freeze-in mechanism \cite{Hall:2009bx} provides a simple alternative to the standard thermal WIMP scenario. In the freeze-in mechanism, the 
DM is very feebly interacting with the cosmic soup, and as a result 
never attains thermal equilibrium in the early universe. 
Hence it is named Feebly Interacting Massive Particles (FIMPs). Due to their very feeble interactions, FIMPs easily escape the direct/indirect detection bounds while satisfying the measured value for the DM relic density (RD). In the present work based on \cite{Abdallah:2019svm}, we explain the above two 
puzzles by extending the SM gauge group by a $U(1)_{B-L}$ gauge symmetry as a simple  extension of the SM.
\begin{figure*}[t!]
\centering
\includegraphics[height=5.0cm,width=7cm]{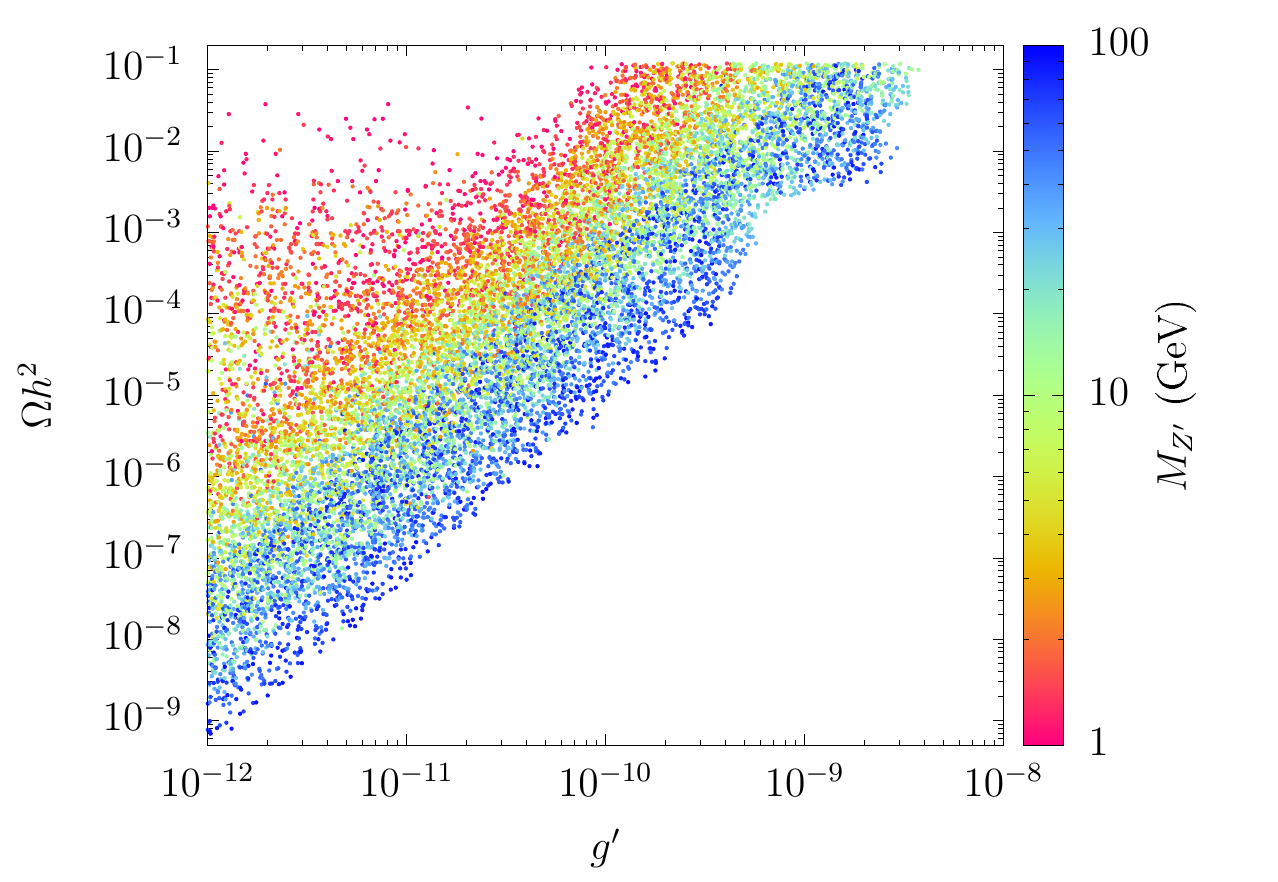}
~~~~\includegraphics[height=5.0cm,width=7cm]{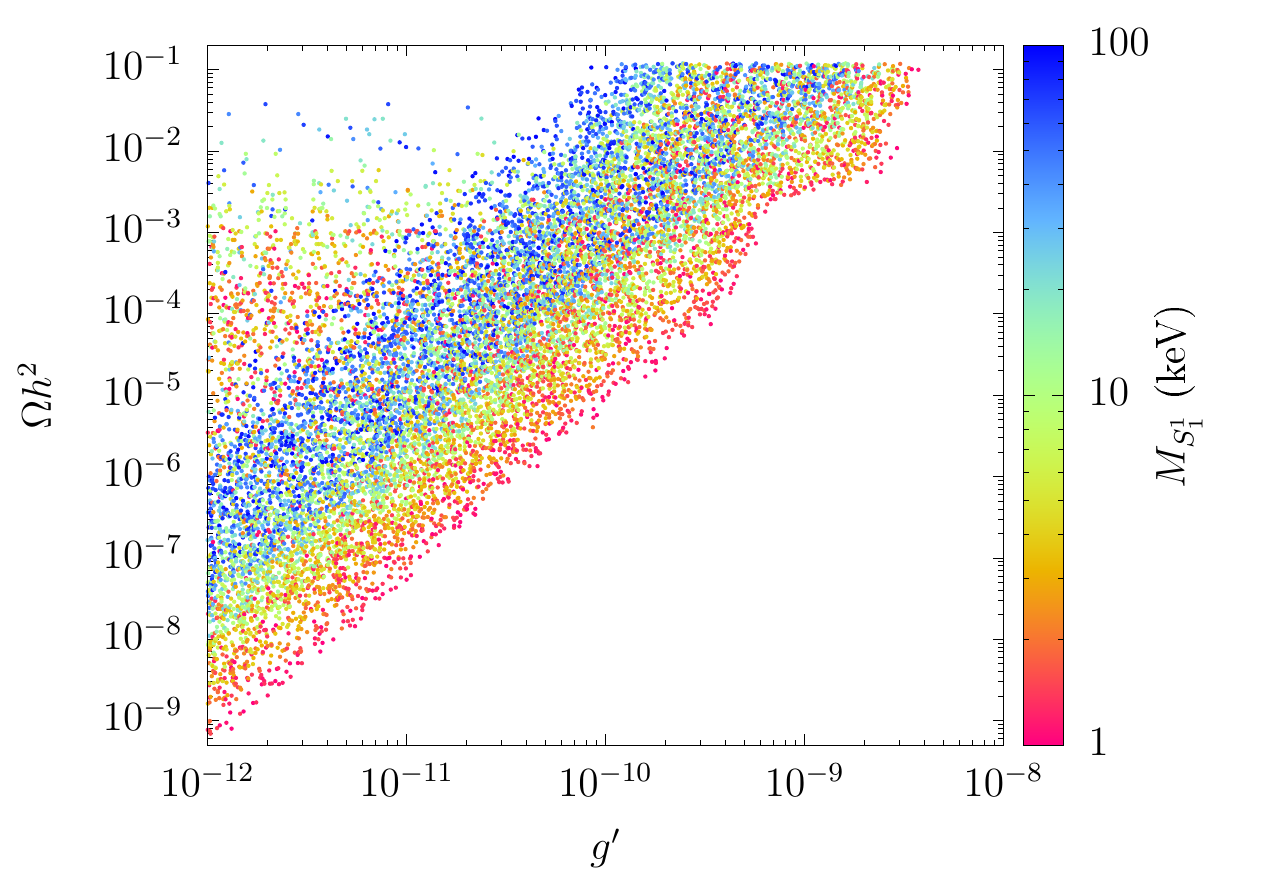}
\caption{Allowed points in $(g^{\prime},\Omega h^2)$
plane after imposing a constraint $\Omega h^2 \leq 0.12$, as an upper bound on the WDM RD, $\Omega h^2$.}
\label{scatt-1}
\end{figure*}
\section{TeV Scale $B-L$ Extension of the SM with Inverse Seesaw (BLSMIS):} 
The $B-L$ extension of the SM is based on the gauge group:
$$SU(3)_C\times SU(2)_L\times U(1)_Y\times U(1)_{B-L}.$$
In this model, nine additional SM singlet fermions ($N_R^i$ and $ S^i_{1,2},~i=1,2,3$) are needed to explain the naturally small neutrino masses through the inverse seesaw mechanism \cite{MohapatraIS1,Mohapatra:1986bd,Khalil:2010iu}.
In addition, an extra neutral gauge boson $Z'$ associated to $U(1)_{B-L}$ and an extra SM singlet scalar, $\phi_H$, are introduced. The full Lagrangian is given by  
\begin{eqnarray}\label{Lag}
\mathcal{L} &=& \mathcal{L}_{\rm SM} 
- \frac{1}{4} F^\prime_{\mu\nu} F^{\prime\,\mu\nu}
+ (D^{\mu} \phi_{H})^{\dagger} D_{\mu} \phi_{H}+\frac{i}{2} \bar{N_{R}} \gamma^{\mu} D_{\mu} N_{R}\nn \\
&+&  \frac{i}{2} \bar{S_{1}} \gamma^{\mu} D_{\mu} S_{1} +\frac{i}{2} \bar{S_{2}} \gamma^{\mu} D_{\mu} S_{2} 
- \mathcal{V}(\phi_h, \phi_{H})\nn \\
&-& (Y_{\nu} \bar{L} \tilde{\phi_h} N_R 
+ Y_S \bar{N}^c_{R} \phi_{H} S_{2} + {\it h.c.}),\nn
\end{eqnarray}
where where $F^\prime_{\mu\nu}$ is the $U(1)_{B-L}$ field strength, $D_{\mu}$ is the covariant derivative, $\tilde{\phi_{h}} = i \sigma_2 \phi_h$ and $\mathcal{V}(\phi_h,\phi_{H})$ is the potential (for more details, see~\cite{Khalil:2010iu}). 
After the $B-L$ and electroweak symmetries breaking and the SM Higgs doublet $\phi_h$ and the SM singlet $\phi_H$ take their vacuum expectation values (vevs), $v$ and $v'$, respectively, the mass matrix of the neutrinos is given by
$${{\cal M}_{\nu}=
\left(%
\begin{array}{cccc}
  0 & M_D & 0&0\\
  M^T_D & 0 & M_N&0 \\
  0 & M^T_N & \mu_S&0\\
  0 &0 & 0&\mu_S\\
\end{array}%
\right)}, $$
where $M_D=\frac{1}{\sqrt{2}}Y_\nu v$ and $ M_N =
\frac{1}{\sqrt 2}Y_{S} v' $. Due to the added $\mathbb{Z}_{2}$ symmetry, $S_1$ is completely decoupled and it only interacts with  $Z'$ with a coupling $g'$. Thus its mass is given as,
\begin{equation}
M_{S_1} = \mu_{S}\,.
\end{equation}
Also, the light and heavy neutrino masses, respectively, are given by
\begin{equation}%
M_{\nu_l} \simeq M_D M_N^{-1} \mu_S (M_N^T)^{-1} M_D^T,~M_{\nu_{H,H'}} \simeq M_N.%
\end{equation}%
One can naturally obtain the light neutrino masses $M_{\nu_l}$ to be of order eV
with $\mu_{S}$ of order keV and $M_N$ of order TeV, keeping Yukawa coupling $Y_\nu$ of order one which leads to interesting signatures at the large hadron collider (LHC)~\cite{Huitu:2008gf}. Therefore, the lightest one, $S^1_1$, will be a stable particle and hence a warm DM (WDM) candidate. Also, the lightest heavy right-handed (RH) neutrino $\nu_H^1$ can be a DM (with mass of order GeV) by tuning its corresponding Yukawa coupling to be very small \cite{Fiorentin:2016avj,DiBari:2016guw}. 
\section{Warm DM as FIMP}
As mentioned above, a WDM $S_1^1$ is produced by the freeze-in mechanism only from its coupling with $Z'$. Therefore, the corresponding gauge coupling $g'$ is taken to be very feeble $\sim\mathcal{O}(10^{-10})$ with the result that $S_1^1$ is never in thermal equilibrium with the cosmic soup. Due to small $g'$, $Z'$ also interacts very feebly with the cosmic soup and never achieves thermal equilibrium, 
\begin{equation}
\frac{\Gamma_{Z'}}{H(T=M_{Z'})}<1,
\end{equation}
where $\Gamma_{Z'}$ is the total decay width of $Z'$ and $H$ is the Hubble parameter. The Boltzmann equation (BE) of $Z'$ distribution function of  is given by~\cite{Konig:2016dzg}
\begin{eqnarray}
\hat{L} f_{Z'} = \sum_{i = 1,\,2} 
\mathcal{C}^{h_i \rightarrow Z' Z'} + 
\mathcal{C}^{Z' \rightarrow\, {\rm all}},
\label{Z-dis-collision}
\end{eqnarray} 
where $f_{Z'}$ is the $Z'$ distribution function, $\mathcal{C}^{h_i \rightarrow Z' Z'}$ is the collision term of $Z'$ production from the decays of scalars $h_{1,2}$
and $\mathcal{C}^{Z' \rightarrow\, {\rm all}}$ is $Z'$ decay collision term (for the expression of these collision terms, see~\cite{Gondolo:1990dk,Edsjo:1997bg}).
Once we get $f_{Z'}$, we then can determine
its co-moving number density by using:
\begin{eqnarray}
Y_{Z'} = \frac{45 \ g }{4 \pi^4 g_{s}(M_{\rm sc}/z_0)} \int_{0}^{\infty}
d\xi_p\, \xi_p^2\, f_{Z'} (\xi_p,z)\,.
\label{number-density}
\end{eqnarray}
The keV DM $S_1^1$ can be produced from $f\bar{f}\to Z'\to S_{1}^1 S_1^1$ (annihilation contribution) and from $Z' \to S_{1}^1 S_1^1$ (decay contribution). 
To determine $Y_{S_1^1}$, we solve the
following BE~\cite{Gondolo:1990dk,Edsjo:1997bg,Iso:2010mv},
\begin{eqnarray}
\frac{d Y_{S_1^1}}{dz}&=&\frac{4 \pi^2}{45} \frac{M_{\rm Pl}\, M_{\rm sc} \sqrt{g_{\star}}}{1.66\, z^2}
\sum_{f}\langle {\sigma v}_{f\bar{f}\rightarrow S_{1}^1 S_1^1} \rangle
\left[\left({Y_{f}^{\rm eq}}\right)^2 -  Y^2_{S_{1}^1} \right]\nn\\
&+&\frac{2\, M_{\rm Pl}\, z\, \sqrt{g_{\star}}}{1.66\, M^2_{\rm sc}\, g_s}\,
\langle \Gamma_{Z' \rightarrow S_{1}^1 S_1^1} \rangle_{\rm NTH} 
\left(Y_{Z'} -  Y_{S_{1}^1} \right).
\label{be-s1}
\end{eqnarray}
The corresponding 
RD of the WDM $S_1^1$ is given by~\cite{Edsjo:1997bg} 
\begin{eqnarray}
\Omega\, h^{2} \simeq 2.755 \times 10^{8}\, \left( M_{S_1^1}/{\rm GeV}\right)\,Y_{S_1^1}(\infty)\,.
\label{relic-density-expressionS1}
\end{eqnarray}
From Fig.~\ref{scatt-1}, it is clearly seen that $\Omega h^{2}$ is inversely proportional to $M_{Z'}$ and directly proportional to $M_{S^1_1}$. More explicitly, for a fixed $g'$ value, larger $\Omega h^{2}$ values correspond to smaller $M_{Z'}$ values (red points) and larger $M_{S^1_1}$ values (blue points). Also, it is worth noting that many points ($\sim 84\%$~of the scanned points) have  a small DM RD ($\Omega h^2\leq 10^{-2}$). Therefore, we discuss in the next section a two component FIMP DM possibility  to get an extra RD contribution from the lightest heavy RH neutrino, $\nu_H^1$, as a GeV scale DM.
\begin{figure*}[t!]
\centering
\includegraphics[height=4.8cm,width=5.6cm]{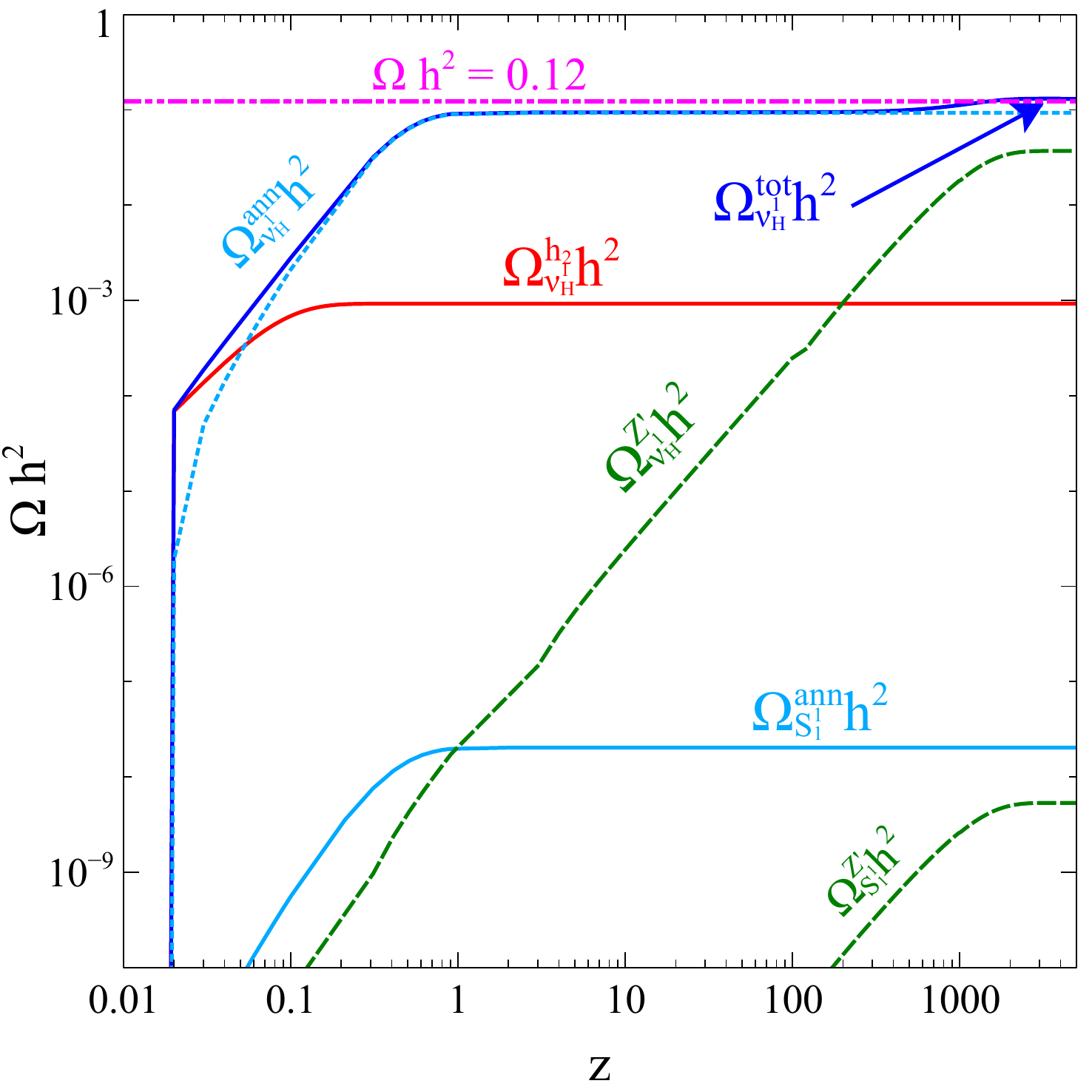}
~\includegraphics[height=4.8cm,width=5.6cm]{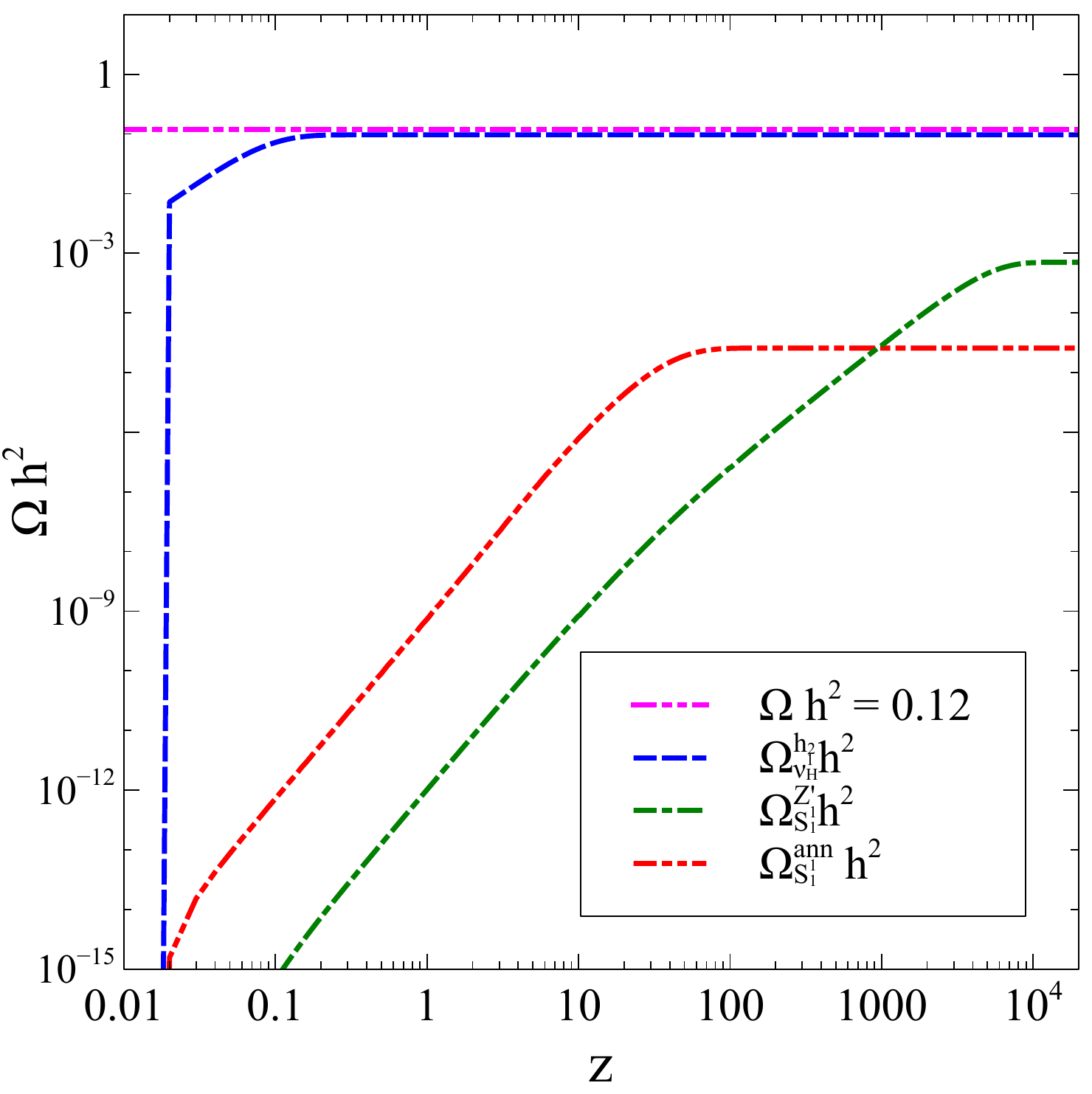}
~\includegraphics[height=4.8cm,width=5.6cm]{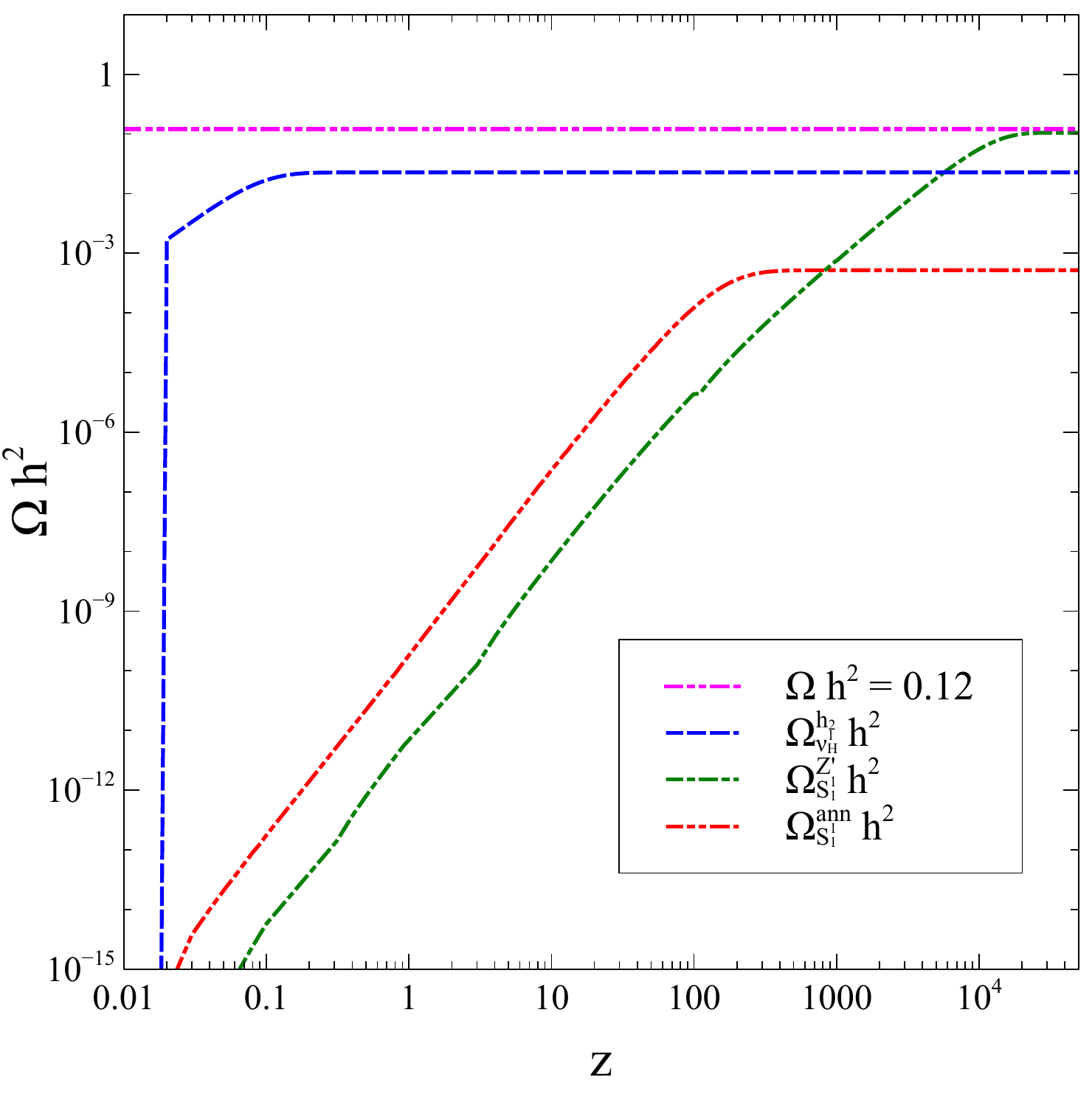}
\caption{Variation of relative RD contributions of $\nu_H^1$ and $S_1^1$ as a function of 
$z$. Here, in left panel: 
$M_{Z'} = 1$~TeV, $M_{\nu_H^1} = 70$~GeV, $M_{S_1^1} = 10$~keV, 
$g' = 9.0 \times 10^{-12}$, $\alpha = 0.01$~rad, and $z_0=0.01$; in center (right) panel:
 $M_{Z'} = 10$~GeV ($2.5$~GeV), $M_{\nu_H^1} = 8$~GeV ($2$~GeV), $M_{S_1^1} = 10$~keV ($100$~keV), $g' \simeq 2.4 \times 10^{-11}$, $M_{h_2} = 5$~TeV, $\alpha = 0.01$~rad, and $z_0=0.01$.}
\label{relic-hi-zp-n1-s1}
\end{figure*}
\section{Two component FIMP DM}
As mentioned, the lightest heavy RH neutrino, $\nu_H^1$, can be a stable  particle by tuning its corresponding Yukawa coupling to be very small $\leq 3\times 10^{-26} ({\rm GeV}/M_N)^{1/2}$~\cite{Fiorentin:2016avj,DiBari:2016guw}. Therefore, it can be an extra DM component, with mass of order GeV. The dominant $\nu^1_H$ pair annihilation channels to SM particles 
 are mediated by the neutral gauge boson $Z'$
and the scalars $h_{1,2}$. The coupling of $\nu^1_H$ pair with $Z'$ is $g'/2$, while with $h_i$ is given by
\begin{eqnarray}\label{HRHRH-coup}
\lambda_{\nu^1_H \nu^1_H h_i} 
= \sqrt{2} \ g' \ \frac{ M_{\nu^1_H} }{M_{Z'}}O_i\,,
\label{n1-n1-h-coupling}
\end{eqnarray}  
where $O_1=\sin\alpha$ and $O_2=\cos\alpha$ ($\alpha$ is the scalar mixing angle). Therefore, $\nu_H^1$ pair annihilation is proportional
to extremly feeble coupling $g'$. Due to this feeble $g'$, $\nu_H^1$ will never reach thermal equilibrium and is  produced by the freeze-in mechanism. The BE associated with $\nu_H^1$ production is as follows~\cite{Gondolo:1990dk,Edsjo:1997bg,Iso:2010mv}
\begin{eqnarray}
\frac{d Y_{\nu^1_H}}{dz} &=&  
\frac{4 \pi^2}{45} \frac{M_{\rm Pl}\, M_{\rm sc} \sqrt{g_{\star}}}{1.66\, z^2}
\sum_{f}\langle {\sigma v}_{f\bar{f}\rightarrow \nu^1_H \nu^1_H} \rangle
\left[\left({Y_{f}^{\rm eq}}\right)^2 -  Y^2_{\nu^1_H} \right]\nn\\
&+&\frac{2\, M_{\rm Pl}\, z\, \sqrt{g_{\star}}}{1.66\, M^2_{\rm sc}\, g_s}\,
\Bigg{[}\langle \Gamma_{Z' \rightarrow \nu^1_H \nu^1_H} \rangle_{\rm NTH} 
\left(Y_{Z'} -  Y_{\nu^1_H} \right) \nn\\
&+&\sum_{i=1,2}
\langle \Gamma_{h_i} \rangle \left(Y^{\rm eq}_{h_i} -  Y_{\nu^1_H} \right)
\Bigg{]}.
\label{be-N1}
\end{eqnarray}
Thermal average of the $h_{1,2}$ decay width is defined as~\cite{Gondolo:1990dk}
\begin{eqnarray}
\langle \Gamma_{h_i} \rangle = \frac{K_{1}(z)}{K_{2}(z)} \Gamma_{h_i}\,,
\end{eqnarray} 
where $\Gamma_{h_i}$ is the total $h_i$ decay width. The corresponding RD
of $\nu^1_H$ is given by~\cite{Edsjo:1997bg}
\begin{eqnarray}
\Omega_{\nu^1_H} h^{2} = 2.755 \times 10^{8}\,
\left(M_{\nu^1_H}/{\rm GeV}\right)\,Y_{\nu^1_H}(\infty)\,.
\label{relic-density-expression}
\end{eqnarray}
Finally, the total
RD of this two component DM is given by
\begin{eqnarray}
\Omega^{\rm tot} h^2 = \Omega_{\nu^1_H} h^{2} + \Omega_{S_1^1} h^{2}\,.
\end{eqnarray}
It is clearly seen that the DM production depends crucially on the mass of the mother particles ($M_{Z'},M_{h_2}$) and the DM mass. 
Assuming $M_{h_2}> 2 M_{Z'} > 4 M_{S_1^1}$, we divide the $\nu_H^1$ spectrum into two regions according to its dominant production modes:
\begin{enumerate}
\item Region~I, where $M_{Z'} > 2 M_{\nu_H^1}$ and $\nu_H^1$ production is $Z'$ dominated,
\item Region~II, where $M_{Z'} < 2 M_{\nu_H^1}$ and $\nu_H^1$ production is $h_2$ dominated.
\end{enumerate}
In region~I, as shown in Fig.~\ref{relic-hi-zp-n1-s1} (left), $\Omega^{Z'}_{\nu_H^1}h^2$ is larger than $\Omega^{h_2}_{\nu_H^1}h^2$ because the latter is suppressed by a factor of their partial decays ratio ($\simeq 12 M^2_{\nu_H^1}M_{h_2}/M^3_{Z'}\simeq {\cal O}(0.1)$). Also, $\Omega_{S_1^1}h^2$  is negligible compared to $\Omega_{\nu_H^1}h^2$ even though they have same gauge coupling $g'$ and their mediator masses ($M_{h_2}$ and $M_{Z'}$) are of the same order ($\sim$~TeV). This is because the RD of a DM candidate is directly proportional to its mass. Therefore, the contribution of the keV mass $S_1^1$ to the DM total RD  as compared to the GeV mass $\nu_H^1$ is suppressed by a factor  $\simeq M_{S_1^1}/M_{\nu_H^1}\simeq {\cal O}(10^{-7})$. In region~II, as shown in Fig.~\ref{relic-hi-zp-n1-s1} (center, right), $Z'$ decays to $\nu_H^1$ pair is kinematically forbidden, and $\nu_H^1$ production consequently is $h_2$ dominated. Therefore, a major portion of the two DM candidates ($\nu_H^1$ and $S_1^1$) is produced almost independently from the $h_2$ and $Z'$ mediated processes, respectively. Moreover, in region~I this possibility did not exist because both $\nu_H^1$ and $S_1^1$ are produced dominantly via $Z'$ and have the same number density.
\section{Conclusion}
We studied two problems beyond the SM, namely, the non-vanishing tiny neutrino masses and the existence of the DM within the BLSMIS.
In the BLSMIS, $S_1^1$ can be a WDM, being odd under a $\mathbb{Z}_2$ symmetry. We studied
$S_1^1$ as a FIMP WDM and found that a large portion of the parameter space gives a small contribution to the DM RD.
Hence, as a possible scenario in the BLSMIS, we considered a two component FIMP DM  to get an extra contribution to the DM RD.  In this scenario, the lightest heavy RH neutrino, $\nu^1_H$, can contribute independently to the DM RD as a GeV scale DM. For $M_{Z'}>2M_{\nu_H^1}$, the production of $\nu_H^1$ through the $Z'$ mediator has the dominant contribution to the total DM RD, while for $M_{Z^{\prime}} < 2 M_{\nu^1_H}$, $h_2$ mediated processes will contribute dominantly to $\nu_H^1$ production and the $Z'$ mediated processes will contribute dominantly to $S_1^1$ production. In this region, we emphasized that both FIMP candidates, $S_1^1$ and $\nu_H^1$, have relevant contributions to the total DM RD.
\section*{Acknowledgements}
The authors would like to thank the Department of Atomic Energy
Neutrino Project of Harish-Chandra
Research Institute (HRI). We also acknowledge the HRI cluster
computing facility (http://www.hri.res.in/cluster/).
This project has received funding from the European Union's Horizon
2020 research and innovation programme InvisiblesPlus RISE
under the Marie Sklodowska-Curie
grant  agreement  No.~690575 and  No.~674896.
\bibliographystyle{unsrt}

\end{document}